\documentclass[fleqn,usenatbib]{mnras}

\usepackage{newtxtext,newtxmath}
\usepackage[T1]{fontenc}

\DeclareRobustCommand{\VAN}[3]{#2}
\let\VANthebibliography\thebibliography
\def\thebibliography{\DeclareRobustCommand{\VAN}[3]{##3}\VANthebibliography}

\usepackage{graphicx}	% Including figure files
\usepackage{amsmath}	% Advanced maths commands
\usepackage{pdflscape}
\title[Metallicities of the NE and W shelves]{The origin of the metallicity distributions of the NE and W stellar shelves \\ in the Andromeda Galaxy}

\author[S. Milo{\v s}evi{\' c} et al.]{
S. Milo{\v s}evi{\' c},$^{1}$\thanks{E-mail: stanislav.milosevic@matf.bg.ac.rs}
M. Mi{\' c}i{\' c},$^{2}$ and
G. F. Lewis$^{3}$
\\
$^{1}$University of Belgrade, Department of Astronomy at Faculty of Mathematics, Belgrade, Serbia \\
$^{2}$Astronomical Observatory of Belgrade, Belgrade, Serbia\\
$^{3}$Sydney Institute for Astronomy, School of Physics, A28, The University of Sydney, NSW 2006, Australia
}

\date{Accepted XXX. Received YYY; in original form ZZZ}

\pubyear{2022}

\begin{document}
\label{firstpage}
\pagerange{\pageref{firstpage}--\pageref{lastpage}}
\maketitle

\begin{abstract}
Tidal streams and stellar shells are naturally formed in galaxy interactions and mergers. The Giant Stellar Stream (GSS), the North-East (NE), and Western (W) stellar shelves observed in Andromeda galaxy (M31) are examples of these structures and were formed through the merger of M31 and a satellite galaxy. Recent observational papers have provided strong evidence that the shells and GSS originate from a single progenitor. In this paper, we investigate the formation of these two stellar shelves and the detailed nature of their relationship to the GSS. We present numerical simulations of tidal disruption of a satellite galaxy assuming that it is a progenitor of the GSS and the shell system. We represent the progenitor as a dwarf spheroidal galaxy with the stellar mass of $10^9$ $M_{\odot}$ and evolve its merger with M31 for 3 Gyrs to reproduce the chemodynamical properties of the NE and W shelves. We find that an initial metallicity of the progenitor with a negative radial gradient of $\Delta$FeH = -$0.3 \pm 0.2$, successfully reproduces observed metallicities of the NE, W shelves, and the GSS, showing that all these structures can originate from the same merger event.
%We use Monte Carlo simulations to describe the final metallicity distributions of the NE and W shelves, and assume a negative radial metallicity gradient, $\Delta$FeH = -$0.3 \pm 0.2$, for the initial distribution in the progenitor galaxy. We calculated the theoretical values of the metallicity for the NE shelf, and these values are between -0.4 and -0.5, and $\sim$ -0.5 is the mean metallicity for the W shelf. With this same gradient, we already reproduced the observed GSS properties, and now observed values for the metallicity of the NE and W shelves. 
\end{abstract}


%%%%%%%%%%%%%%%%%%%%%%%%%%%%%%%%%%%%%%%%%%%%%%%%%%

%%%%%%%%%%%%%%%%% BODY OF PAPER %%%%%%%%%%%%%%%%%%

%\begin{figure*}
%\centering
%\includegraphics[scale=0.8]{Figure1.png}
%\caption{Orientation of the GSS in x-y plane after 2.7 Gyrs. The centre of the image is in the centre of the disk of M31 and M31 is not shown for clarity. Left panel shows all of the particles from the dwarf progenitor, while the right panel shows the matching density plot. Over-plotted in the left panel are eight observational fields.}
%\end{figure*}

\section{Introduction}
The Andromeda Galaxy (M31) is our cosmic neighbour and due to its proximity, it is very important for studying galactic dynamics and evolution. In the hierarchical assembly paradigm of galaxy formation, large mass galaxies, like Andromeda, are formed through mergers of smaller galaxies (\citealp{White1, White2}). Some of the tracers of these mergers are structures strewn through the halos of massive galaxies such as streams and shells (\citealp{Hernquist1988, Johnston2001, Johnston2008}). 

Many observed structures in the halo of M31  are formed due to merger events of M31 with satellite galaxies. The most prominent is Giant Stellar Steam (GSS) discovered by \citet{Ibata2001}.  Distances along the GSS were given by \citet{McConnachie2003}, \citet{Conn}, and velocities by \citet{Ibata2004}, \citet{Guhathakurta}, \citet{Kalirai2006a}, \citet{Gilbert2009}. The stellar mass of the stream is $\approx 2.4 \times 10^{8}M_{\odot}$ (\citealp{Ibata2001, Fardal2006}). In the work of \citet{Ferguson2002}, the discovery of the North Eastern and Western shelves is presented. By comparing color-magnitude diagrams these two structures probably have the same origin as GSS (\citealp{Ferguson2005, Richardson2008}). Further investigations were done in \citet{Merret2006} and \citet{Bhattacharya} where observations of the planetary nebulae in the region of the NE shelf are presented. Using observations from \citet{Ferguson2002}, in the work of \citet{Fardal2007} the W shelf is detected. One faint structure, in the southeast, the SE shelf is detected spectroscopically in \citet{Gilbert2007}. The observations that unveil the kinematics of RGB stars in the W shelf are given in \citet{Fardal2012}. The observational fields of the edges of the shells are given in \citet{Fardal2008}. The edge of the NE shelf is placed on the radial distance from the center of M31 of $\approx$ 40 kpc, and the edge of the W shelf at $\approx$ 20 kpc. The consistency of the physical properties based on color-magnitude diagrams and photometric metallicities arises the probability of the same progenitor for these tidal structures (\citealp{Ferguson2005, Gilbert2007, Richardson2008, Tanaka2010, Fardal2012, Bernard}). The evidence that tidal structures in M31 halo are related to the GSS is given in \citet{Brown2006, Brown2008}.

In general, the formation of the structures in the halo of the host is discussed in \citet{Pillepich}, \citet{Remus}, \citet{Karademir}, \citet{Milosevic2022}. The formation of shells is typical for radial mergers \citep{Amorisco}. In the halo of M31, the origin of the GSS and NE and W shelves are discussed in theoretical works of \citet{Fardal2006, Fardal2007, Fardal2013} \citet{Sadoun}, \citet{Hammer2010, Hammer2013}. In the work of \citet{Fardal2007}, it was shown in numerical simulations that the GSS and shelves could have the same progenitor, confirmed in follow-up work by \citet{Sadoun}. In \citet{Fardal2006,Fardal2007} satellite on very radial orbit forms these substructures in the halo of M31. In our previous paper, \citet[][hereafter Paper I]{Milosevic},  we also showed that the GSS, the NE, and W stellar shells can all have the same progenitor. The stellar mass of the satellite in \citet{Fardal2006}, \citet{Sadoun} and Paper I is the same $\sim$ $10^{9}M_{\odot}$, but in \citet{Sadoun} and Paper I there is a 20 times more massive halo of the progenitor galaxy. These minor merger scenarios reproduced the stream and shell system. Unlike minor merger scenarios, the major merger (stellar mass of the progenitor $\sim$ $10^{10}M_{\odot}$) was presented in several works (\citealp{Hammer2018, DB}) with less success in reproducing properties of the shelves than minor merger scenarios. 

The morphology of the progenitor is still an open question. A disky progenitor was used in models of \citet{Fardal2008}, \citet{Kirihara2017}, \citet{MikiMori} and dwarf spheroidal galaxy (dSph) in the \citet{Sadoun}. In Paper I, we also used the dSph model for the progenitor of the GSS and shelves. Both models successfully reproduce these structures, although disky progenitor better reproduces the asymmetric structure of the GSS envelope. In these theoretical works, structures are formed from the same progenitor. Despite the morphology of the progenitor, the NE shelf forms after the second and the W shelf after the third pericentric passage. Recent observational works show strong evidence that the stream and shells have a common origin based on their dynamics (\citealp{Escala2022, Dey}).

Metallicity values in the region of the halo substructures of M31 were presented in the Elemental Abundances in M31 survey (\citealp{Gilbert2019, Escala2020a, Escala2020b, Escala2021}). The halo of M31 is also observed in the SPLASH survey, where kinematics and metallicities are given (\citealp{Kalirai2006a, Gilbert2007, Gilbert2009, Fardal2012, Wojno}). For the GSS are given metallicities along (\citealp{Conn, Cohen}) and across the stream (\citealp{Guhathakurta, Kalirai2006a, Ibata2007, Gilbert2009, Gilbert2014}). The observed values give two gradients along the stream, where metallicity values increase from -0.7 in the inner part of the GSS to the central part where the value is -0.2, and then in the outer part, metallicity drops off at a value around -0.8. In the direction across the stream, there is a gradient between metal rich core and the envelope of the stream.

In the region of the NE and W shelves, are given metallicities and velocities based on RGB stars observations for several spectroscopic fields. Due to spectroscopic observations, in \citet{Fardal2012} are given velocities and metallicities for several fields in the W shelf. These fields are between 13 and 26 kpc of the projected radius. In the phase space, projected radius vs. line-of-sight velocity ($R_{\rm proj}$-$v_{\rm los}$) the observed sample shows a wedge pattern which is expected for the shells, formed in the almost radial mergers. It is clearly detected in \citet{Fardal2012} and suggested that the W shelf is formed in the third pericentric passage of the same progenitor that formed the GSS in the first passage. The first analysis of the metallicity and kinematics of the RGB stars in the NE shelf is given in \citet{Escala2022}. The observed fields for the NE shelf are from 13 to 31 kpc of projected radius. The wedge pattern for the NE shelf is detected supporting that this structure is formed in radial merger in the second pericentric passage. The observed metallicity of the NE shelf is [Fe/H]$_{\rm phot}$ = -0.42, and for the W shelf is [Fe/H]$_{\rm phot}$ = -0.55. The shapes of MDFs are similar between shells and stream supporting the common origin. Without a kinematically  observed trace of the core of the progenitor in \citet{Escala2022}, it is given more arguments for models with fully disrupted progenitor.

The observed metallicity gradients along and across the GSS unlocked the possibility of testing initial metallicity distributions in the progenitor of the stream (\citealp{Fardal2008, MoriRich, Fardal2013, MikiMori, Kirihara2017}).
In Paper I, we modeled the distribution of the metallicity in the progenitor of the GSS. Radial metallicity gradients in dwarf galaxies can be positive and negative (\citealp{Koleva2009a,Koleva2009b,Spolaor}) and we showed that negative radial metallicity gradient in progenitor can reproduce final metallicity distribution along and across the GSS. In the theoretical work of \citet{Mercado}, from FIRE2 simulation, it is also given that dSph galaxies in the Local Group have a linearly decreasing metallicity gradient. Results are compared with observed metallicity profiles from \citet{Leaman, Kacharov}. Here we calculate the theoretical metallicity in the regions of the NE and W shelves as it is likely that these structures were formed in the same merger event as the GSS. At the same time with known metallicity values it is possible to predict the position of the remnant of the progenitor. Matching the observed metallicity in several fields in the NE shelf, with the theoretical one from the initial distribution in the progenitor, we have an additional test for the initial gradient and connection between the shell system and the GSS.  

This paper is organized as follows: In Section 2 we introduce the method and N-body models for M31 and the satellite progenitor of the GSS. Also, we present the MC method for finding the initial metallicity distribution in the dwarf progenitor galaxy. In Section 3 we present the main results based on the comparison of simulations to observations. In Section 4 we discuss our results and conclude.
 
\section{Methods}

\subsection{N-body models}

In the following, we use a N-body model for M31 and the progenitor galaxy. These models are described in detail in Paper I and are similar as given in \citet{Geehan2006}, \citet{Fardal2007}, and \citet{Sadoun}. For the main morphological parts of M31, we assume a disk, bulge, and dark matter halo and for the progenitor, we assume a spherical baryonic part and dark matter halo. The initial conditions were generated with GalactICs package \citep{Widrow}, which computes the positions and velocities of the particles for a given mass model. We ran our simulation with Gadget2 cosmological TreePM code \citep{Springel}. We assume a total mass of M31 to be $\sim 10^{12} M_{\odot}$. \\
\indent The bulge is represented with the Prugniel-Simien profile \citep{Widrow}, which is a de-projected Sersic profile. This profile has $r^{1/n}$ law:
\begin{equation}
\rho_{b}=\rho_{b0} \left( \frac{r}{r_{b}} \right) \textrm{exp} \left( r/r_{b} \right)^{-1/n}.
\end{equation}
Here, $\rho_{b0}$ is the density at $r=r_{b}$, and $r_{b}$ is a spherical scale radius for the bulge, and the value for $n$ is 1.8.   

The disk is represented by a combination of two profiles: exponential profile of surface density in the x-y plane and $\rm sech^{2}$ law in the vertical, z-direction.  
The exponential profile is given with (\citealp{Geehan2006, Sadoun}):
\begin{equation}
\Sigma(R)=\frac{M_{d}}{2\pi R_{d}^{2}} e^{-\frac{R}{R_{d}}}.
\end{equation}
\noindent Here, $M_{d}$ is the total mass of the disk, $\Sigma$ is surface density, and $R_{d}$ is disk scale radius. In the last two equations, $r$ is the spherical radius, and $R$ is the cylindrical radius.

The $\rm sech^{2}$ profile is used in vertical (z) direction \citep{Sadoun} and combined profile is given by:
\begin{equation}
\rho(R,z)=\frac{\Sigma(R)}{2z_{0}} \textrm{sech}^{2} \left( \frac{z}{z_{0}} \right).
\end{equation}
\noindent Here, $z_{0}$ is the scale height of the disk. The inclination of the disk is $77^{o}$ and the position angle is $37^{o}$ \citep{Fardal2007}, and the heliocentric distance to Andromeda is taken to be 785kpc. 

A spherical dark matter halo is represented with Navarro-Frenk-White profile \citep{Navarro1996}:
\begin{equation}
\rho_{h}(r)=\frac{\rho_{0}}{\frac{r}{r_{s}}\left(1+\frac{r}{r_{s}}\right)^{2}}
\end{equation}
\noindent where $\rho_{0} = \sigma_{h}^{2}/4\pi r_{s}^{2}$ is a characteristic density, and $\sigma_{h}^{2}$ is a characteristic
velocity dispersion, $r_{s}$ is the scale radius. A more general form is given in GalactICS \citep{Widrow} with an additional term for truncation at some point.
\begin{equation}
\rho(r)=\frac{2^{2-\alpha}\sigma_{h}^{2}}{4\pi r_{s}^{2}}\frac{1}{(r/r_{s})^{\alpha}(1+r/r_{s})^{3-\alpha}}\frac{1}{2} \textrm{erfc} \bigg{(}\frac{r-r_{h}}{\sqrt{2} \delta_{r_{h}}}\bigg{)}.
\end{equation}
\noindent Here, $r_{h}$ is the radius of the halo and the value at which density starts to decrease, $\delta_{r_{h}}$ is the distance along which density falls to zero, $\alpha$ is an exponent in NFW profile, and we took $\alpha = 1$. %The tidal radius $r_{200}$ is taken to be approximately equal to the distance where density drops 200 times in comparison to central density.

The total mass of the halo inside $r_{200}$ radius is $M_{200}=8.8\times 10^{11}M_{\odot}$. The values of parameters are used in GalactICS in the way described in \citet{Widrow}, where N-body models of M31 are presented, to generate initial conditions. Parameters of the M31 galaxy and dwarf galaxy are given in Table 1 and Table 2 respectively. 

\begin{table*}
\centering 
\begin{center}
\begin{tabular}{|c|c|c|c|c|c|c|c|} 
\hline
\hline
component & m [$M_{\odot}$] & N \\
\hline
Bulge &  $3.36 \times 10^{5}$  & 96247 & $r_{b}$ = 1,23 kpc & $\sigma_{b}=$ 393 km/s & $M_{b}= 3.2 \times 10^{10} M_{\odot}$   \\
\hline
Disk &  $3.36 \times 10^{5}$ & 108929 & $R_{d}$ = 6.82 & $z_{0}$= 0.57 & $M_{d}=$ $3.66 \times 10^{10} M_{\odot}$\\
\hline
Halo & $3.36 \times 10^{6}$ & 261905 & $r_{h}$= 122.5 kpc & $r_{s}=$ 8 kpc & $M_{h} = 8.8 \times 10^{11} M_{\odot}$ & $\delta_{r_{h}}$= 12 kpc & $\sigma_{h}$ = 525 km/s  \\
\hline
\end{tabular}
\end{center}
\caption{The values of the parameters for the N-body model of M31 used in GalactICS, same as in Paper I, where m is the mass of one particle, and N is the number of particles in each component. Similar values are used in \citet{Geehan}, \citet{Fardal2007}, and \citet{Sadoun}.}
\end{table*}
%with parameter values motivated from observational works of \citet{Kent}, \citet{Braun} and \citet{Widrow2003}.

\begin{table*}
        \label{tab:landscape}
        \begin{tabular}{lccccr} 
            \hline
            \hline
            component & m [$M_{\odot}$] & N  \\
            \hline
            Baryonic matter & $1.66 \times 10^{4}$ & 131072 & $r_{s}=$ 1.03 kpc & $\sigma_{s}=$ 93 km/s & $M_{s}=2.18 \times 10^{9} M_{\odot}$ \\
            \hline
            Dark matter halo & $1.66 \times 10^{5}$ & 248809 & $r_{h}=$ 5 kpc & $\sigma_{h}=$ 242 km/s & $M_{h}=4.13 \times 10^{10}M_{\odot}$\\
             \hline
        \end{tabular}
        \caption{The values of the parameters for the N-body model of progenitor galaxy, same as in Paper I, where m is the mass of one particle, and N is the number of particles in each component. Similar values for the baryonic matter are used in \citet{Sadoun}.}
\end{table*}

We assume that the best timescale for comparison of modeled and observed properties is between 2 and 3 Gyrs (Paper I). That is in agreement with previous results (\citealp{Sadoun, Hammer2018}). The dwarf spheroidal galaxy starts its very radial orbit with null velocity from a distance of around 200kpc from the center of the M31 galaxy. The dynamical history of the formation of the shells is the same as in the case of the formation of the GSS because we use the same model and the same conditions in N-body simulation as in previous work.

\subsection{Metallicity distribution}

We used the same model of metallicity distribution in the progenitor galaxy as described in Paper I, assuming a negative gradient from the center of the galaxy in the radial direction. We successfully reproduced metallicity distribution in the GSS with gradient $\Delta$FeH = -$0.3 \pm 0.2$ and a central metallicity value of -0.2, matching the observed metallicity values for the GSS taken from the \citet{Conn} and \citet{Cohen}. Many theoretical models of the formation of the GSS predict the formation of shells after several orbits (\citealp{Fardal2008, Kirihara2017, MikiMori}). Accordingly, we try with the same model to describe the observed photometric metallicity of the NE and W shelves given in \citet{Escala2022}.

We used Monte Carlo (MC) methods to probe the initial metallicity distribution in the progenitor galaxy. The galaxy is divided into spherical shells and for each particle in every shell, we randomly attach the metallicity value from the Gaussian distribution. The central value of the distribution in each shell is picked up from the negative radial metallicity gradient. We also took $\sigma = 0.4$ for the Gaussian distribution of metallicity in the shell. We trace these particles through simulation and calculate final metallicity values in spectroscopic fields from \citet{Escala2022}. After 1000 iterations we calculate the mean value and standard deviation of metallicity in each observed field to generate the final distribution.

We took values from 0 to -0.3 in the center of the progenitor and from -1 to -1.8 in the outer part as described in Paper I. These values and metallicity gradient $\Delta$FeH = -$0.3 \pm 0.2$ we used to describe the final metallicity distribution in the GSS. This distribution was motivated by observed metallicity along the GSS, given in \citet{Conn}. With fixed all parameters except initial distribution in the progenitor, we investigated the final distribution in the NE and W shelves and compared it with observed values given in the spectroscopic fields from \cite{Escala2022}.

\section{Results}
In the works of \citet{Fardal2008} and \citet{Sadoun}, and in Paper I, the formation of the shells is represented as the same merger event in which the GSS is formed. Here we present the formation of the NE and W shelves in a similar single merger scenario with modeled metallicity values for these shelves. Also, we analyze the complex kinematical structure of the shell system. 

\subsection{Formation and kinematics of the NE and W shelves}
On almost radial orbit, a satellite of the M31 galaxy experiences tidal disruption and forms the GSS and shell system. In the first pericentric passage, the GSS is formed, in the second the NE shelf, and in the third the W shelf. The NE shelf is formed in front of the M31 galaxy, closer to us, and the GSS and the W shelf are further from us \citep{Ferguson2005}.
We presented the formation of the GSS and shelves in Figure 1. Only particles from the progenitor galaxy are presented and M31 particles are omitted for the sake of clarity. We can see formed GSS and observed fields from \citet{McConnachie2003} are given in red crosses and from \citet{Conn} in blue dots. Green dots are the edges of the shelves given in \citet{Fardal2008}, and black dots observed fields from \citet{Fardal2012} for the W shelf and \citet{Escala2022} for the NE shelf. For these fields, we calculated metallicity from our simulation. The shells presented in coordinates as we see in the sky are not as prominent as the GSS. In the \citet{Hammer2018} the timescale for the merger event and formation of the structures in the halo of M31 is between 2 and 3 Gyrs; in \citet{Sadoun} was suggested 2.7 Gyrs for the GSS, and we estimated (Paper I) the best timescale for the formation of the GSS, from the metallicity distribution point of view, which is 2.9 Gyrs.  
 We can see more clearly formed shells in the middle panel, where the surface density plot is given in the $\xi$-$\eta$ angular coordinates. On the right panel in Figure 1 are presented shells and the stream in the y-z plane, where we can see that particles of the NE shelf are formed closer to us, with negative values of z-coordinate, unlike the W shelf that is further away from us. 

\begin{figure*}
\centering
\includegraphics[scale=0.85]{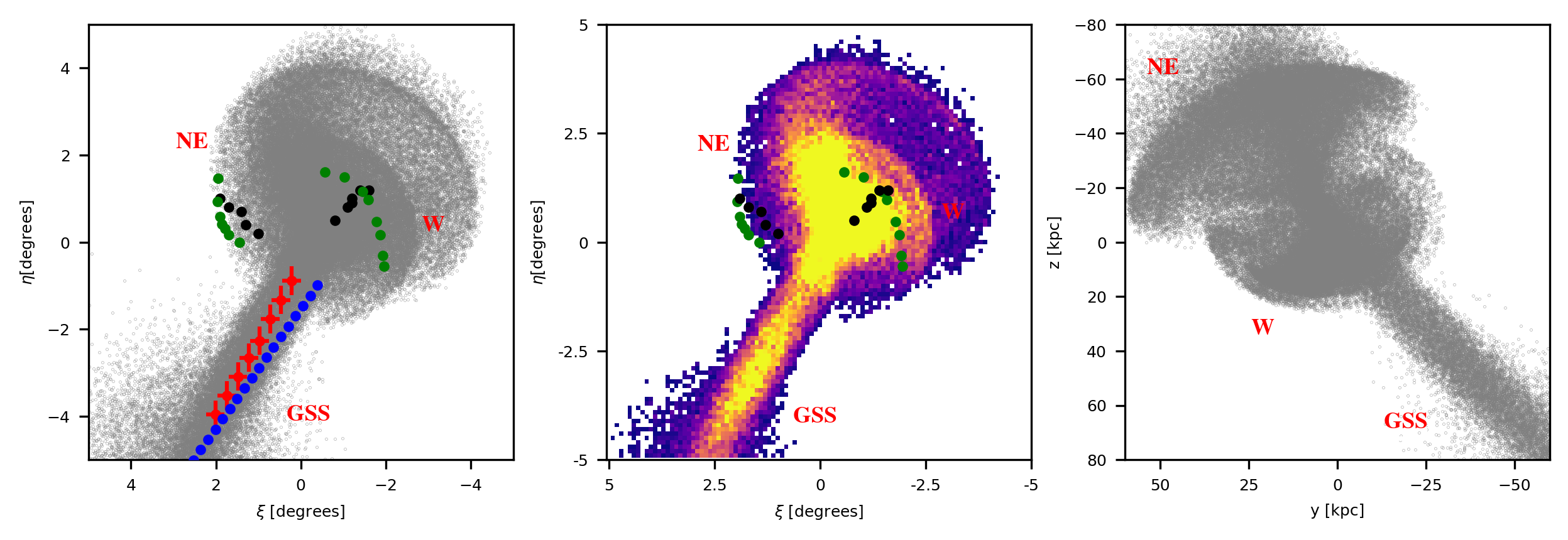}
\caption{The formation of the GSS, NE, and W shelves after 2.4 Gyrs from the beginning of the simulation. Red crosses are the observed fields given in \citet{McConnachie2003}, blue dots in \citet{Conn}, and green dots are edges of the NE and W shelves given in \citet{Fardal2008}. Observed fields for the NE shelf from \citet{Escala2022} are given in black dots and also for the W shelf from \citet{Fardal2012}.}
\end{figure*}

\begin{figure}
\centering
\includegraphics[scale=0.52]{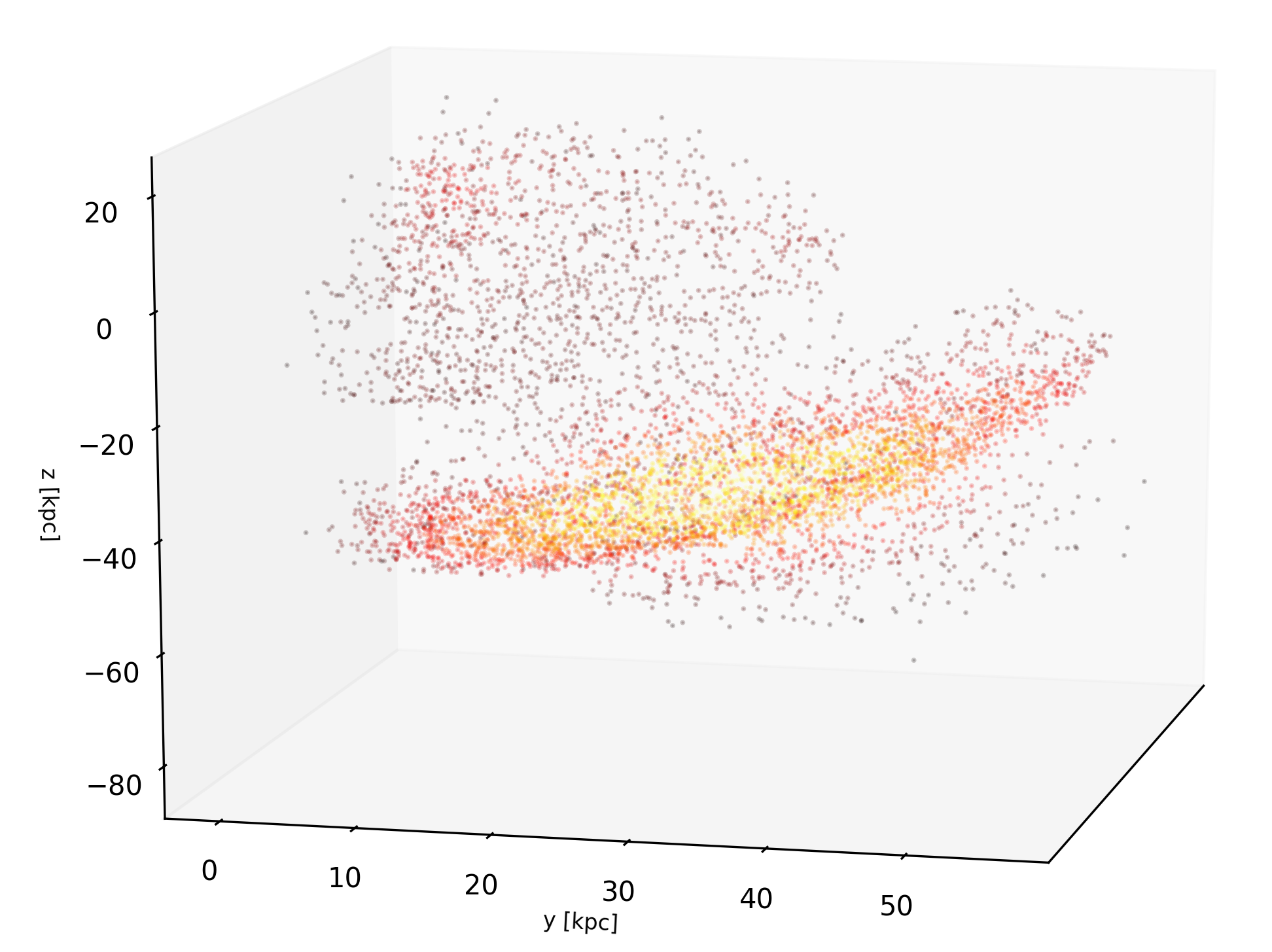}
\caption{The 3D plot of the NE shelf at 2.4 Gyrs from the beginning of the simulation. The z-coordinate is in the line of sight. The coordinate system is centered on M31. Color represents particle density.}
\end{figure}

\begin{figure}
\centering
\includegraphics[scale=0.54]{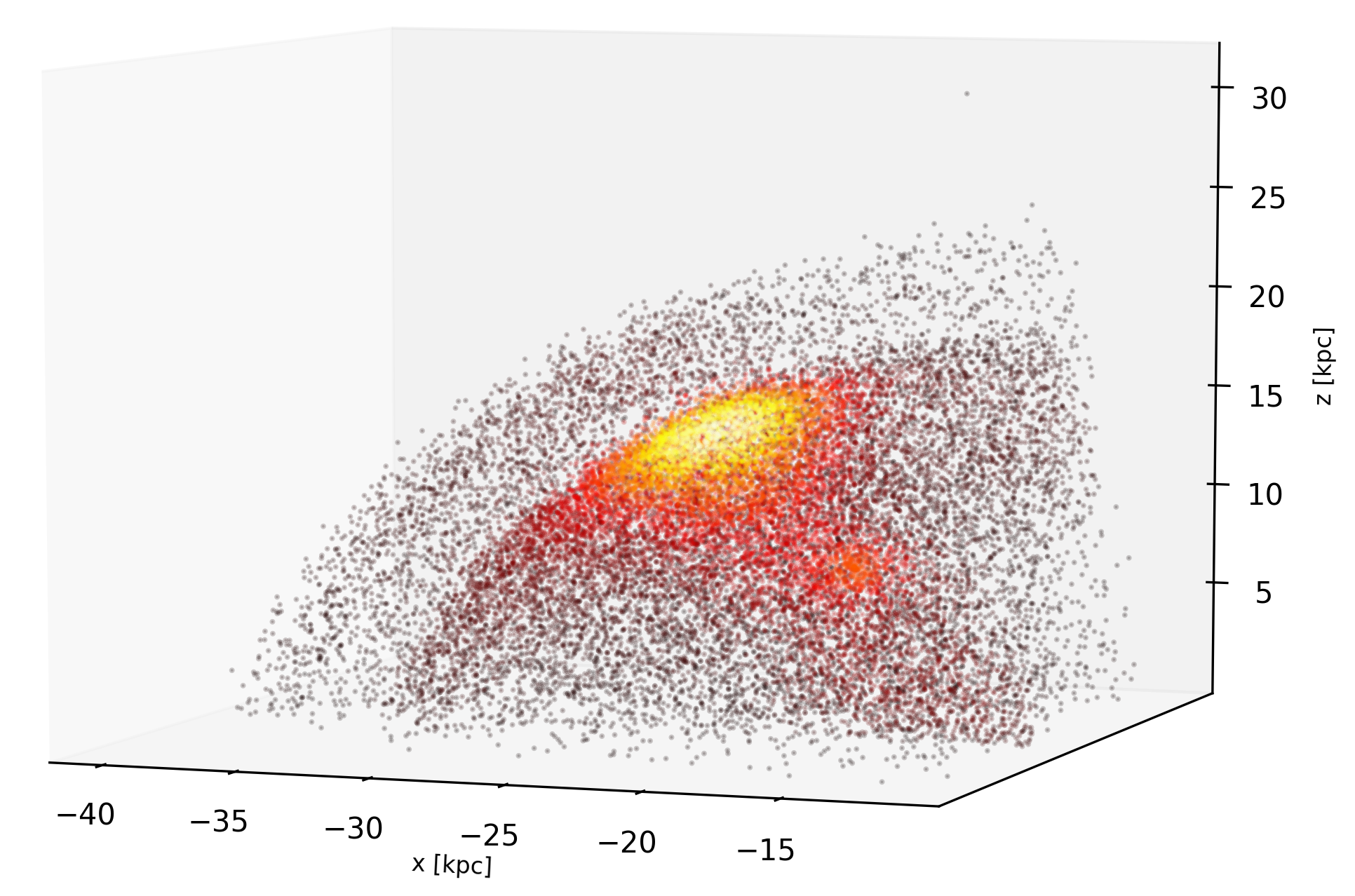}
\caption{The 3D plot of W shelf at 2.4 Gyrs from the beginning of the simulation. The coordinate system is the same as in Figure 2. Color represents particle density.}
\end{figure}

We took the particles from the regions of the NE and W shelves and represented them in 3D space in Figure 2 and Figure 3. In Figure 3 are given particles of the NE shelf. We can see the edge of the structure. As suggested in \citet{Fardal2008}, the NE shelf is formed closer to us, due to the M31 galaxy. The particles in our simulation will have negative values for the z-coordinate. Unlike the NE shelf, the W shelf will form further away from us, due to M31, in the region of the positive z-coordinate. We cannot detect the remnant of the progenitor in this extracted tidal structure in 3D space. 

Structures, such as streams and shells, are better presented in phase-space plots, where the x-axis is given the distance from the center of M31, and the y-axis is given radial velocity relative to M31. In Figure 4 are given phase-space plots for the time interval between 2.3 and 2.8 Gyrs. We can see the formed GSS and the NE and W stellar shelves, and also the remnant of the progenitor is presented in the vicinity of the NE shelf. Although there is no observational evidence for the remnant in \citet{Escala2022}, we can see it in phase-space plots clearly, and after 2.6 Gyrs it becomes faint. The reason for that could be complete disruption.

%In Paper I, we estimated the timescale of 2.9 Gyrs for the best match between simulated and observed data, and from Figure 5 as well as from previous figures, the best timescale for the shell system is 2.5 Gyrs.   

\begin{figure*}
\centering
\includegraphics[scale=0.95]{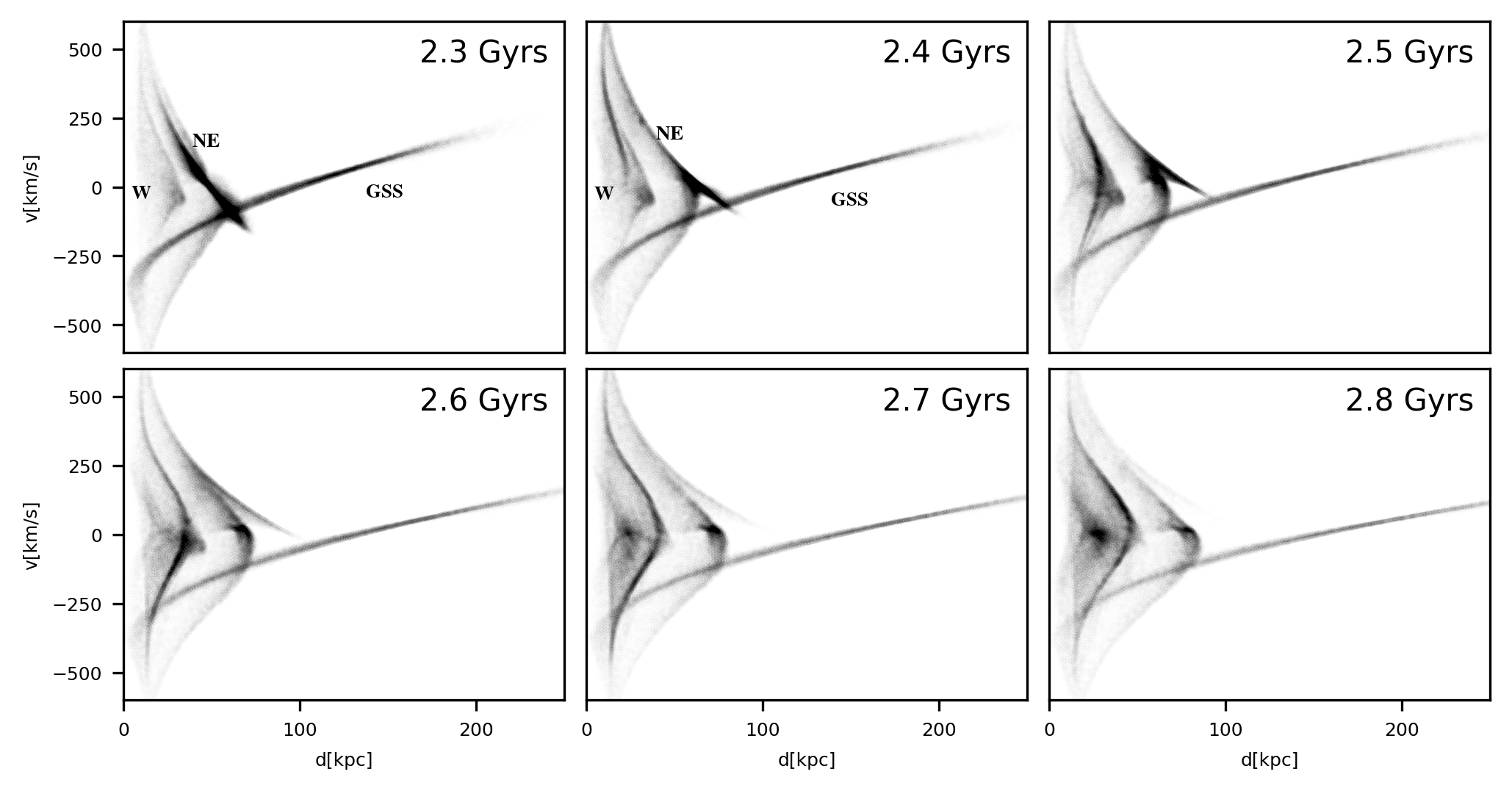}
\caption{The phase-space plots in $d-v_{r}$ for different timescales between 2.3 and 2.8 Gyrs. We can see the evolution of the tidal structures: GSS, NE, and W shelves on this time interval.}
\end{figure*}

For the shell system, we can see a characteristic wedge pattern in the space of projected radius ($R_{ \rm proj}$) vs. $v_{\rm los}$. In Figure 5 evolution of this pattern for the NE and W shelves, and the evolution of the GSS over a time interval between 2.3 and 2.8 Gyrs are presented. From \citet{Fardal2008}, \citet{Fardal2012} and \citet{Escala2022} the estimated $R_{\rm proj}$ for the NE shelf is $\approx$ 40kpc and for the W shelf is $\approx$ 20 kpc, although some observed fields are beyond these distances. At the moment of 2.4 Gyrs, the tip of the wedge is at 40kpc; after that, the NE shelf is spreading further. The best time scale for forming the NE shelf, based on the shape of the characteristic wedge pattern is 2.4 Gyrs. 

%From these plots, we can see that the distance of the outer edge of the shells is in agreement with estimated values from \citet{Sadoun} and \citet{Escala2022} for the timescale of 2.5 Gyrs. 

In  Paper I, we presented the very complex kinematic nature of the GSS, where two main flows are responsible for the observed peaks in the metallicity distribution in the stream. In the works of \citet{Fardal2007, Fardal2013}, a kinematic analysis of the shell system in the halo of M31 and the distribution of the observed stars over velocities were given. Spectroscopic measurements of the W shelf are given in \citet{Fardal2012} which lead to the kinematic structure of the shell as well as metallicity values for the target objects in several fields. \citet{Escala2022} made additional observations of the NE shelf. From these observations, it is found two Kinemacically Cold Components (KCC) in the region of the NE shelf. Similar to the GSS, this shell also has a complex kinematic structure. 

To compare with observed data from \citet{Escala2022} we took particles from the region of the NE shelf. The NE shelf region is defined with boundaries $X_{M31}>0.5^{\circ}$ and $-1.5^{\circ}<Y_{M31}<0.5^{\circ}$, where $X_{M31}$ is NE major axis (position angle = $38^{\circ}$ E of N), and $Y_{M31}$ NW minor axis \citep{Escala2022}. In Figure 6 (left panel), the structure shows a wedge shape with the tip of the wedge at 40 kpc. Blue rectangles represent spectroscopic fields from \citet{Escala2022}. On the right panel of Figure 6 are given particles from the W shelf with spectroscopic fields also represented with blue rectangles. The part of the NE shelf with positive line-of-sight velocities ($v_{\rm los}$), $v_{\rm helio}-v_{\rm M31}>0$ (where systemic velocity of M31 is $v_{\rm M31}$ = -300km/s), the upper envelope, corresponds to the stars moving toward M31, and the lower envelope corresponds to stars that are moving in the opposite direction \citep{Escala2022}. This shows a very complex kinematical picture of the NE shelf. We calculated mean velocities and velocity dispersions in our simulation that correspond to spectroscopic fields given in \citet{Escala2022} and compared them with the observed one in Figure 7. The upper envelope is represented with red color and the lower envelope with blue color. Smaller dots represent our model, and larger dots observed values of $v_{\rm los}$. In most of the fields, we can see agreement between modeled and observed values, but there is also mismatching in the first field ($R_{\rm proj}$=14.4 kpc), where the model failed to explain values for the lower envelope, as well as in the third field ($R_{\rm proj}$=21.7 kpc). 

A similar analysis is done for the W shelf. In Figure 8 we see a comparison between observed and modeled mean velocities and velocity dispersions for the upper and lower envelopes in the W shelf. The upper envelope is represented with red color and the lower with blue. Larger dots are observed values and small dots form our simulation. There is an agreement between these values in all spectroscopic fields. 

%To compare with observed data from \citet{Escala2022} we made the distribution of the particles over velocities and we picked up particles from our simulation from the regions of the NE shelf represented in Figure 6 with blue rectangles. These regions are observed regions in \citet{Escala2022} and dots are particles from the N-body simulation. The distributions over velocities are shown in Figure 7, where two rectangles represent observed velocity intervals for the two KCCs. The comparison is made for the timescale of 2.9 Gyrs from the beginning of the simulation. For other time intervals, an agreement is not as good as for this one. The only discrepancy between the simulated distribution and observed kinematic structure is for the KCC1 in field NE6. In the fields NE2 and NE4 from \citet{Escala2022}, KCC1 and KCC2 are overlapped.  

\begin{figure*}
\centering
\includegraphics[scale=0.95]{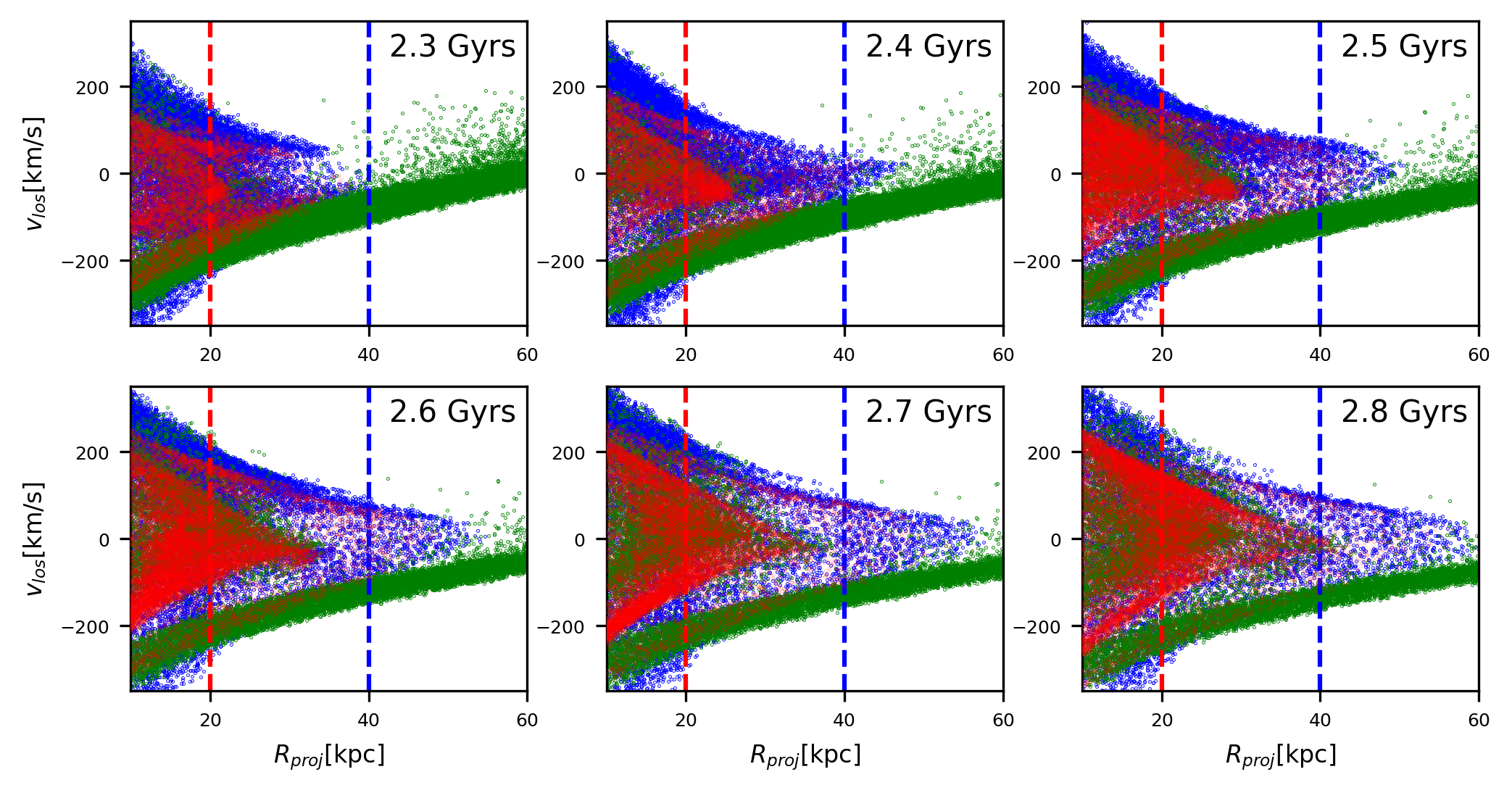}
\caption{The phase-space density plot in $R_{\rm proj}-v_{\rm los}$ plane from 2.3 to 2.8 Gyrs. With green color are presented particles from the GSS; with blue the NE shelf, and with red the W shelf. Red and blue dashed lines present suggested positions from observations \citealp{Fardal2007, Escala2022} of the tip of the wedge pattern for the W and NE shelves, respectively.}

\end{figure*}

\begin{figure*}
\centering
\includegraphics[scale=0.9]{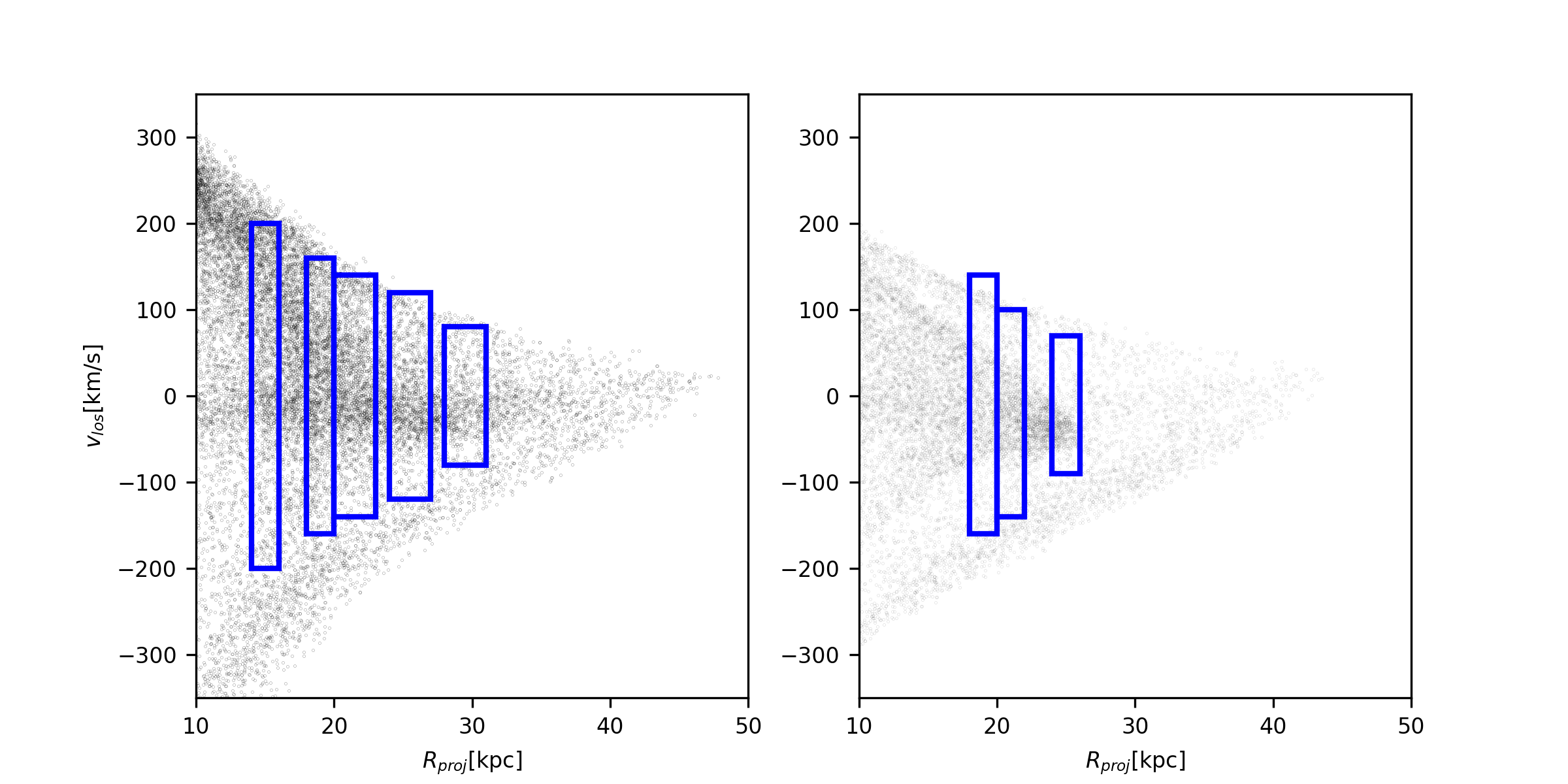}
\caption{The phase-space density plot in $R_{\rm proj}-v_{\rm los}$ at 2.4 Gyrs. Blue rectangles are observed regions in the NE shelf (left panel) and the W shelf (right panel) from \citet{Escala2022}.}
\end{figure*}

\begin{figure}
\centering
\includegraphics[scale=0.5]{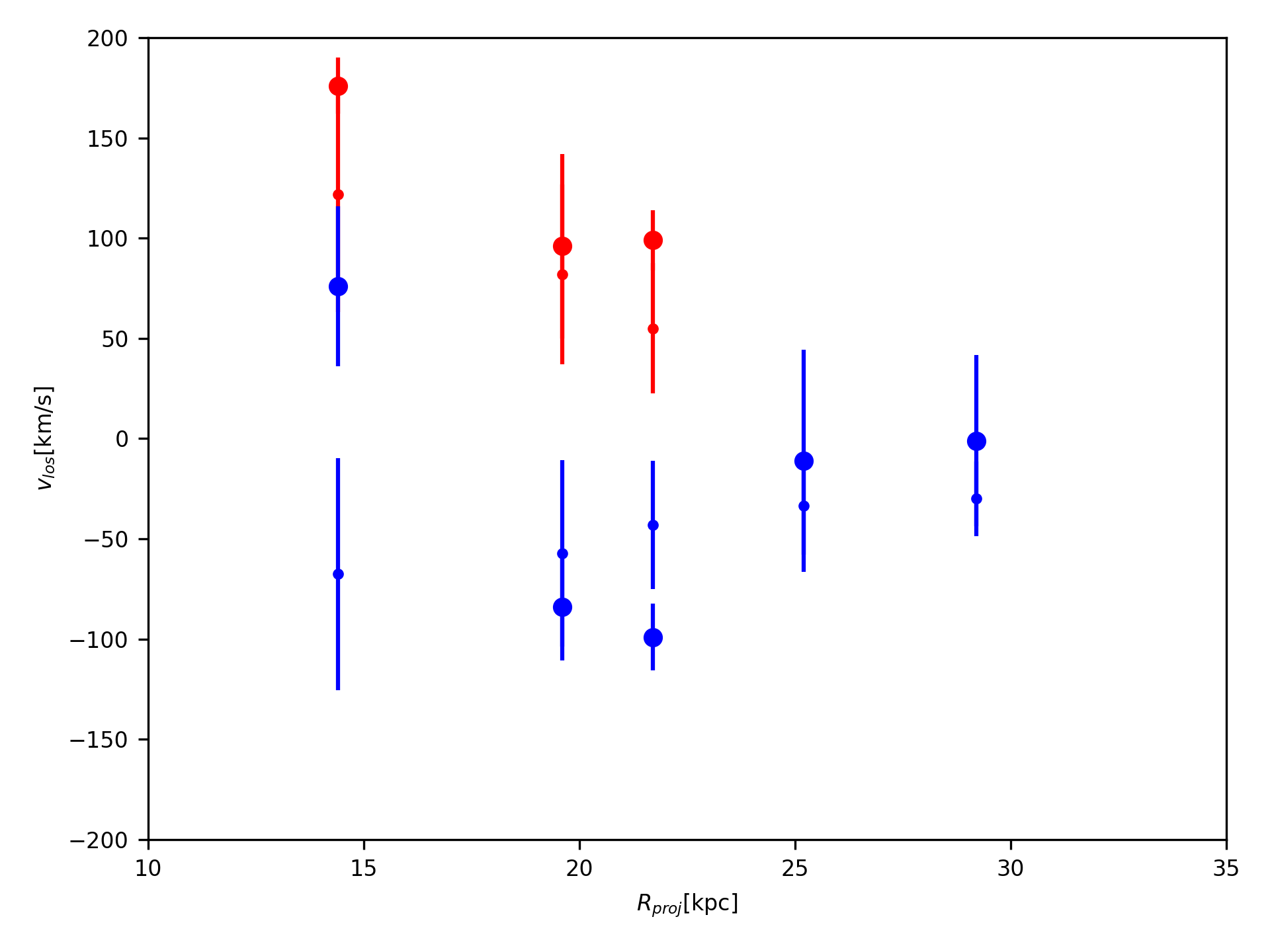}
\caption{The comparison between simulated (small dots) and observed (large dots) mean velocities in the upper (red) and lower (blue) envelope in the NE shelf, at 2.4 Gyrs. The observed values of the mean velocities and velocity dispersions are taken from \citet{Escala2022}.}
\end{figure}

\begin{figure}
\centering
\includegraphics[scale=0.5]{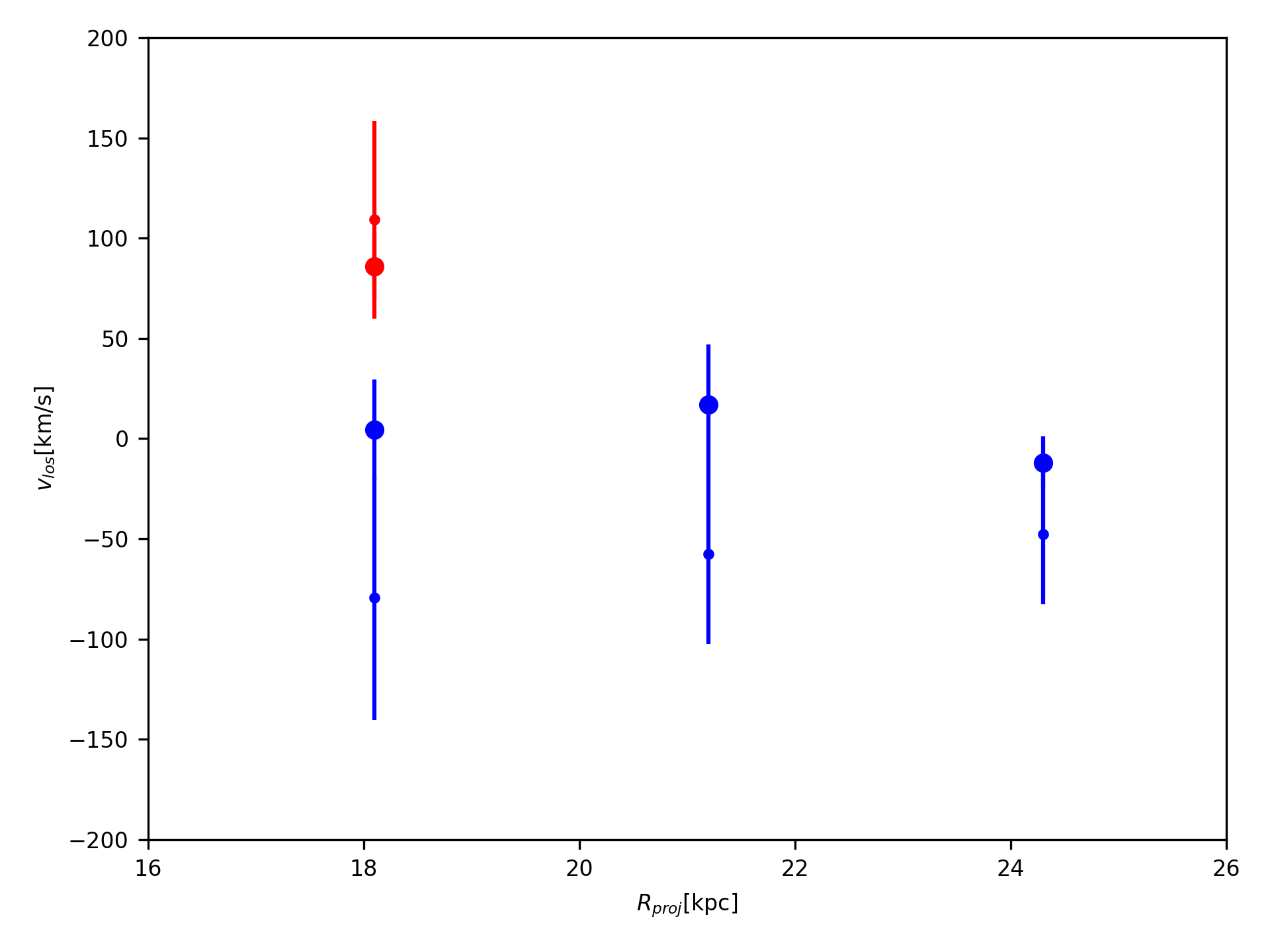}
\caption{The comparison between simulated (small dots) and observed (large dots) mean velocities in the upper (red) and lower (blue) envelope in the W shelf, at 2.4 Gyrs. The observed values of the mean velocities and velocity dispersions are taken from \citet{Escala2022}.}
\end{figure}

 %For the W shelf, we compare the velocity distribution from the simulation with the observed interval in Figure 8. We took all particles from the region of the W shelf and analyze the evolution of the velocity distribution. It is found that most particles have radial velocities of around 0 km/s or heliocentric velocity around -300 km/s. The simulated velocity distribution is in agreement with the observed one. The spatial distribution of the particles and kinematic structure of both shelves and the GSS supports the hypothesis of the common origin.

\subsection{Metallicity distribution of the NE and W shelves}

In Paper I, we showed how the initial metallicity distribution can explain observed metallicity in the GSS. Observed values were given in the works of \citet{Conn} and \cite{Cohen}. These values are distributed along the GSS, and earlier were published the values of the metallicities across the GSS in several fields (\citealp{Ibata2007, Gilbert2007, Gilbert2009}). The similar metallicity values of the GSS and the NE and W shelves support the scenario of the formation of these structures in the same merger event. From the metallicity distribution functions (MDF) for the GSS, NE, and W shelves presented in \citet{Escala2022}, we can see the distributions are very similar for these three structures. Observed metallicity values are given in five fields for the NE and three radial zones for the W shelf. The observed metallicity values in the GSS and shelves and similarity in the MDFs gave the possibility of probing initial metallicity distribution in the progenitor galaxy. 

We investigate the negative metallicity gradient in the progenitor galaxy, the same one we used in Paper I to compare with observed metallicity in \citet{Escala2022}. We calculated metallicity in 5 fields for the NE shelf and also in 5 fields for the W shelf. The good agreement is shown in Figure 9 for the NE shelf. Between 2.4 and 2.7 Gyrs we reproduced observed metallicity in all fields except for the first field, where theoretical metallicity is higher than the observed one, probably because of the contamination of the first field by the M31 stars. Based on these fields we cannot see any obvious gradient in the NE shelf as it is clear for the GSS. All values calculated from the model are very close to the mean value of -0.4 given from observations.

\begin{figure*}
\centering
\includegraphics[scale=0.9]{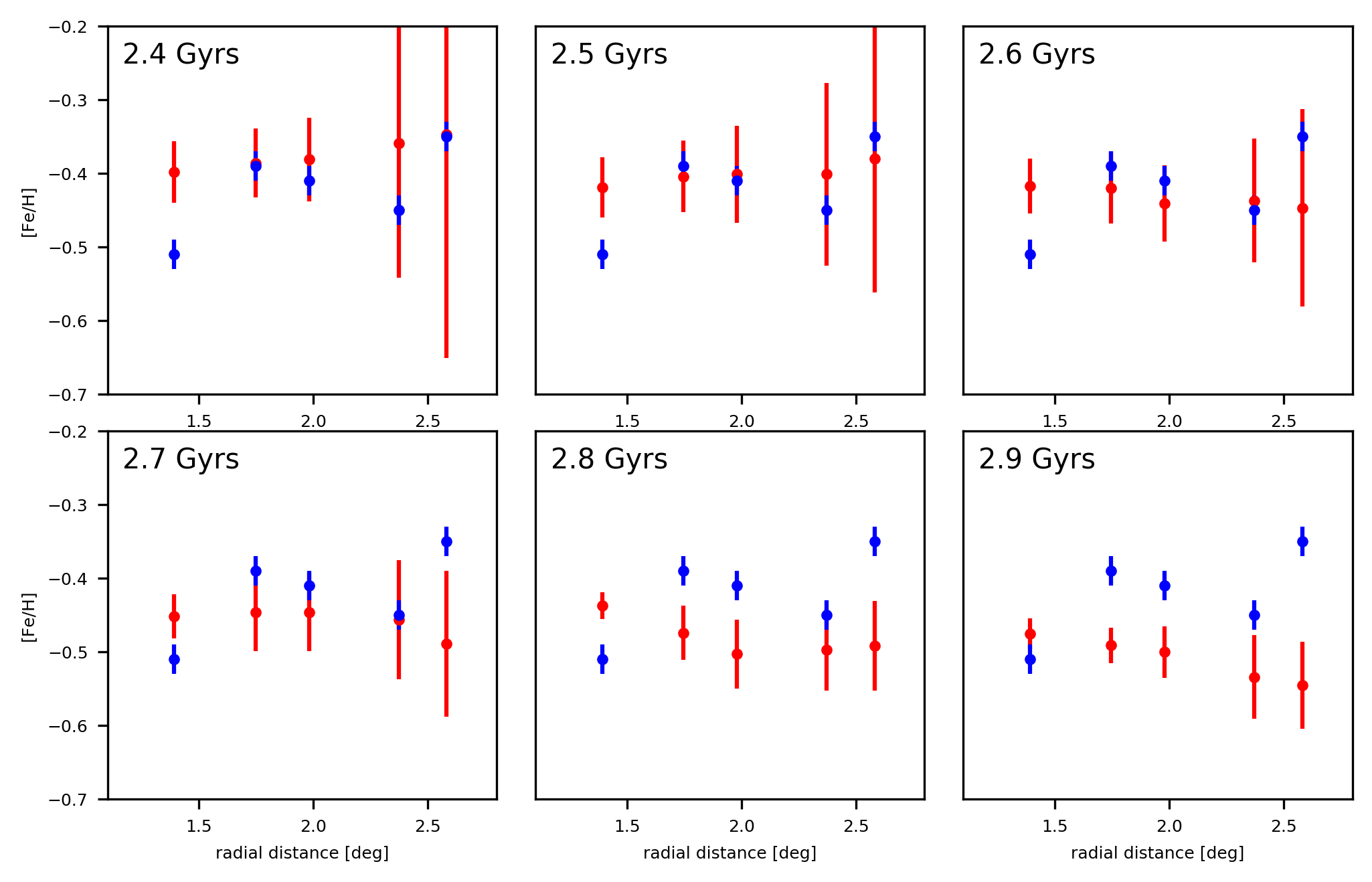}
\caption{Metallicity distribution for the observed fields in the NE shelf. Blue dots are observations from \citet{Escala2022} and red dots are simulated metallicities.}
\end{figure*}

\begin{figure*}
\centering
\includegraphics[scale=0.9]{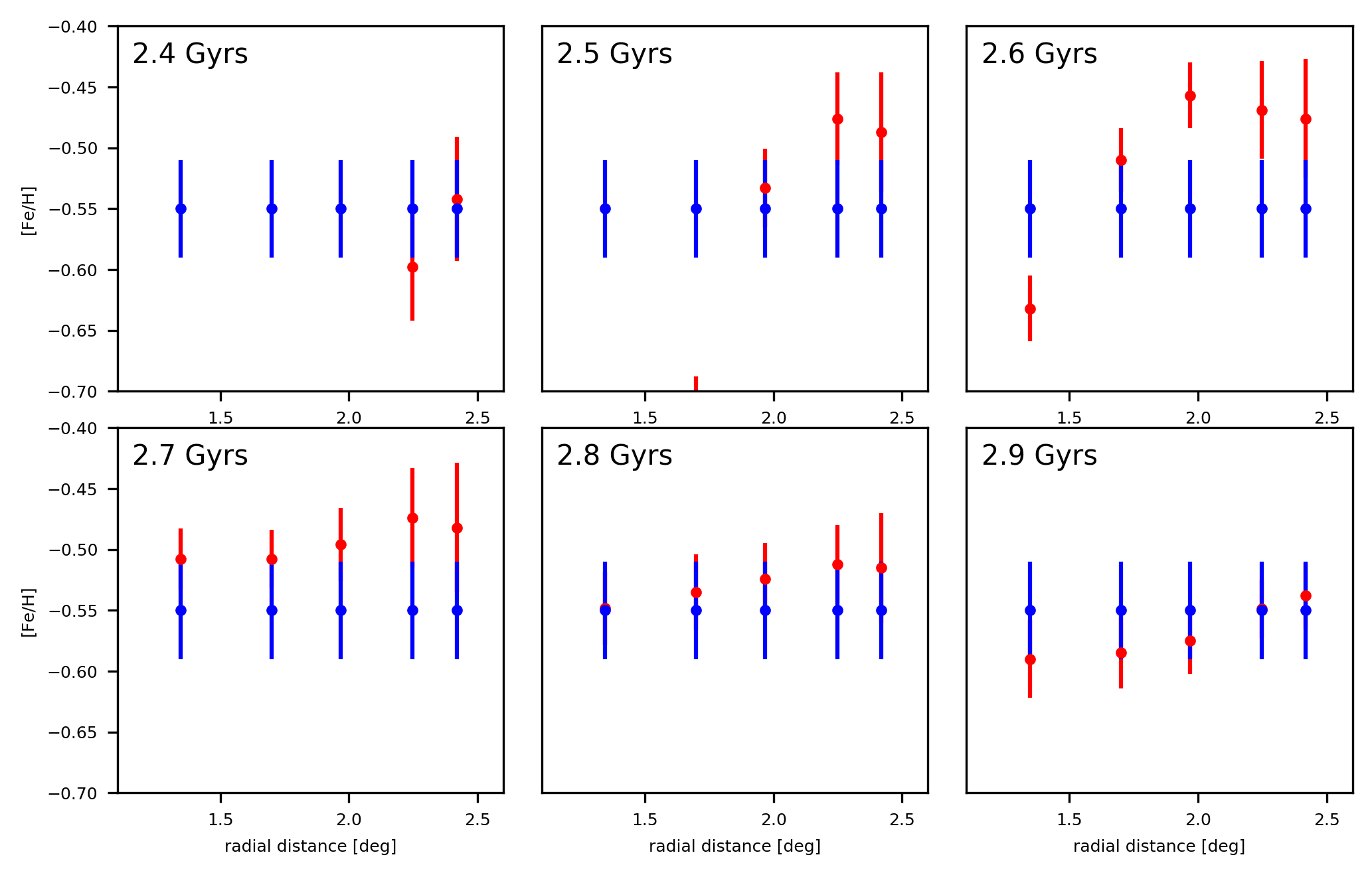}
\caption{Metallicity distribution for the observed fields in the W shelf. Blue dots are observations from \citet{Escala2022} and red dots are simulated metallicities.}
\end{figure*}

We compare our theoretical metallicity for the W shelf in spectroscopic fields with the mean values given in \citet{Escala2022}, so in each field is given mean metallicity for the W shelf from \cite{Escala2022} instead of the observed value, just for the qualitative comparison. These results are presented in Figure 10. The W shelf is more metal-poor than the NE shelf with mean metallicity $\sim$ -0.5.  Similar to the NE shelf, we cannot see a gradient in the W shelf. Between 2.7 and 2.9 Gyrs, there is also good agreement between our model and observed metallicity values, suggesting the same progenitor for all three structures. 

\section{Discussion and conclusions}

This paper is a continuation of Paper I, where we investigated metallicity distribution in the GSS. In many theoretical models, these shelves are formed in the same merger scenario as the GSS (e.g. \citealp{Fardal2008, Sadoun}).

Here, we reproduced the formation of the NE and W shelves using N-body simulations. We used dSph progenitor with a stellar mass of $10^{9}M_{\odot}$. Based on phase-space plots we found that the best agreement with the observed wedge shape we have for 2.4 Gyrs after the beginning of the simulation.  We compared simulated values of mean velocities in the upper and lower envelopes in the NE shelf with observed one from several spectroscopic fields, given in \citet{Escala2022}. Our model does not reproduce the values of the mean velocities in the first field at the projected radius of 14.4 kpc, and in the third field at the projected radius of 21.7 kpc for the upper envelope in the NE shelf. In the three other fields, there is an agreement between simulated and observed values for velocities. For the W shelf, our model successfully reproduces observed values from \citet{Escala2022}.

We used a negative radial metallicity gradient in the dSph progenitor. In Paper I, with radial metallicity gradient $\Delta$FeH = -$0.3 \pm 0.2$ and with a central value of -0.2 we reproduced metallicity distribution along the GSS given in \citet{Conn}. The first observations and analysis for the W shelf were performed by \citet{Fardal2012}. We compared our theoretical metallicity with the mean metallicity for the W shelf given in \citet{Escala2022}, for the same spectroscopic fields as in \citet{Fardal2012}. The simulated metallicity is in agreement with the observed one, except for the innermost field for the NE shelf, where disagreement is probably due to M31 contamination. The theoretical and observed values for the mean metallicity for the NE shelf is $\sim - 0.4$ and for the W shelf, $\sim - 0.5$. From different criteria, the timescale for the formation of the GSS and shell system is between 2.4 and 2.9 Gyrs. 

The disruption of the remnant of the progenitor is still an open question and the morphology of the progenitor is as well. We traced the remnant of the progenitor, to 3 Gyrs, which is presented in phase-space plots. After 2.7 Gyrs into the merger, it is hard to locate the remnant. The suggestion from the \citet{Escala2022} is that the remnant is fully disrupted, or it could be located in the region of the disk of M31 (\citealp{Fardal2013}), which leads to difficulties in observing. Here, we avoided the static potential for the M31 galaxy and used N-body models for both: progenitor and M31. This can affect tidal disruption and dynamic friction as well as the location of the remnant. Major merger scenarios have different predictions for the remnant of the progenitor, a compact dense galaxy like M32 \citep{DB}, or complete disruption \citep{Hammer2018}.

Our model of the spheroidal progenitor successfully reproduced the observed metallicity of the GSS, represented in the Paper I and NE and W shelves represented in this paper. With a negative radial gradient of initial metallicity in the progenitor, we described metallicity distribution in the GSS and NE shelf and reproduced mean metallicity in the W shelf.

\section*{ ACKNOWLEDGEMENTS} 
We thank the anonymous referee for providing important and useful comments that improved this manuscript. This work was supported by the Ministry of Education, and the Ministry of Science, Technological Development and Innovation of the Republic of Serbia, through contracts: 451-03-47/2023-01/200104 and 451-03-47/2023-01/200002.

\section*{Data Availability}
Data used in the paper will be made available on a reasonable request to the author. 



\bibliographystyle{mnras}
\bibliography{NEW} % if your bibtex file is called example.bib

\begin{thebibliography}{}
\makeatletter
\relax
\def\mn@urlcharsother{\let\do\@makeother \do\$\do\&\do\#\do\^\do\_\do\%\do\~}
\def\mn@doi{\begingroup\mn@urlcharsother \@ifnextchar [ {\mn@doi@}
  {\mn@doi@[]}}
\def\mn@doi@[#1]#2{\def\@tempa{#1}\ifx\@tempa\@empty \href
  {http://dx.doi.org/#2} {doi:#2}\else \href {http://dx.doi.org/#2} {#1}\fi
  \endgroup}
\def\mn@eprint#1#2{\mn@eprint@#1:#2::\@nil}
\def\mn@eprint@arXiv#1{\href {http://arxiv.org/abs/#1} {{\tt arXiv:#1}}}
\def\mn@eprint@dblp#1{\href {http://dblp.uni-trier.de/rec/bibtex/#1.xml}
  {dblp:#1}}
\def\mn@eprint@#1:#2:#3:#4\@nil{\def\@tempa {#1}\def\@tempb {#2}\def\@tempc
  {#3}\ifx \@tempc \@empty \let \@tempc \@tempb \let \@tempb \@tempa \fi \ifx
  \@tempb \@empty \def\@tempb {arXiv}\fi \@ifundefined
  {mn@eprint@\@tempb}{\@tempb:\@tempc}{\expandafter \expandafter \csname
  mn@eprint@\@tempb\endcsname \expandafter{\@tempc}}}

\bibitem[\protect\citeauthoryear{{Amorisco}}{{Amorisco}}{2015}]{Amorisco}
{Amorisco} N.~C.,  2015, \mn@doi [\mnras] {10.1093/mnras/stv648}, \href
  {https://ui.adsabs.harvard.edu/abs/2015MNRAS.450..575A} {450, 575}

\bibitem[\protect\citeauthoryear{{Bernard} et~al.,}{{Bernard}
  et~al.}{2015}]{Bernard}
{Bernard} E.~J.,  et~al., 2015, \mn@doi [\mnras] {10.1093/mnras/stu2309}, \href
  {https://ui.adsabs.harvard.edu/abs/2015MNRAS.446.2789B} {446, 2789}

\bibitem[\protect\citeauthoryear{{Bhattacharya}, {Arnaboldi}, {Gerhard},
  {McConnachie}, {Caldwell}, {Hartke}  \& {Freeman}}{{Bhattacharya}
  et~al.}{2021}]{Bhattacharya}
{Bhattacharya} S.,  {Arnaboldi} M.,  {Gerhard} O.,  {McConnachie} A.,
  {Caldwell} N.,  {Hartke} J.,   {Freeman} K.~C.,  2021, \mn@doi [\aap]
  {10.1051/0004-6361/202038366}, \href
  {https://ui.adsabs.harvard.edu/abs/2021A&A...647A.130B} {647, A130}

\bibitem[\protect\citeauthoryear{{Brown}, {Smith}, {Ferguson}, {Rich},
  {Guhathakurta}, {Renzini}, {Sweigart}  \& {Kimble}}{{Brown}
  et~al.}{2006}]{Brown2006}
{Brown} T.~M.,  {Smith} E.,  {Ferguson} H.~C.,  {Rich} R.~M.,  {Guhathakurta}
  P.,  {Renzini} A.,  {Sweigart} A.~V.,   {Kimble} R.~A.,  2006, \mn@doi [\apj]
  {10.1086/508015}, \href
  {https://ui.adsabs.harvard.edu/abs/2006ApJ...652..323B} {652, 323}

\bibitem[\protect\citeauthoryear{{Brown} et~al.,}{{Brown}
  et~al.}{2008}]{Brown2008}
{Brown} T.~M.,  et~al., 2008, \mn@doi [\apjl] {10.1086/592686}, \href
  {https://ui.adsabs.harvard.edu/abs/2008ApJ...685L.121B} {685, L121}

\bibitem[\protect\citeauthoryear{{Cohen} et~al.,}{{Cohen} et~al.}{2018}]{Cohen}
{Cohen} R.~E.,  et~al., 2018, \mn@doi [\aj] {10.3847/1538-3881/aae52d}, \href
  {https://ui.adsabs.harvard.edu/abs/2018AJ....156..230C} {156, 230}

\bibitem[\protect\citeauthoryear{{Conn} et~al.,}{{Conn} et~al.}{2016}]{Conn}
{Conn} A.~R.,  et~al., 2016, \mn@doi [\mnras] {10.1093/mnras/stw513}, \href
  {https://ui.adsabs.harvard.edu/abs/2016MNRAS.458.3282C} {458, 3282}

\bibitem[\protect\citeauthoryear{{D'Souza} \& {Bell}}{{D'Souza} \&
  {Bell}}{2018}]{DB}
{D'Souza} R.,  {Bell} E.~F.,  2018, \mn@doi [Nature Astronomy]
  {10.1038/s41550-018-0533-x}, \href
  {https://ui.adsabs.harvard.edu/abs/2018NatAs...2..737D} {2, 737}

\bibitem[\protect\citeauthoryear{{Dey} et~al.,}{{Dey} et~al.}{2023}]{Dey}
{Dey} A.,  et~al., 2023, \mn@doi [\apj] {10.3847/1538-4357/aca5f8}, \href
  {https://ui.adsabs.harvard.edu/abs/2023ApJ...944....1D} {944, 1}

\bibitem[\protect\citeauthoryear{{Escala}, {Gilbert}, {Kirby}, {Wojno},
  {Cunningham}  \& {Guhathakurta}}{{Escala} et~al.}{2020a}]{Escala2020a}
{Escala} I.,  {Gilbert} K.~M.,  {Kirby} E.~N.,  {Wojno} J.,  {Cunningham}
  E.~C.,   {Guhathakurta} P.,  2020a, \mn@doi [\apj]
  {10.3847/1538-4357/ab6659}, \href
  {https://ui.adsabs.harvard.edu/abs/2020ApJ...889..177E} {889, 177}

\bibitem[\protect\citeauthoryear{{Escala}, {Kirby}, {Gilbert}, {Wojno},
  {Cunningham}  \& {Guhathakurta}}{{Escala} et~al.}{2020b}]{Escala2020b}
{Escala} I.,  {Kirby} E.~N.,  {Gilbert} K.~M.,  {Wojno} J.,  {Cunningham}
  E.~C.,   {Guhathakurta} P.,  2020b, \mn@doi [\apj]
  {10.3847/1538-4357/abb474}, \href
  {https://ui.adsabs.harvard.edu/abs/2020ApJ...902...51E} {902, 51}

\bibitem[\protect\citeauthoryear{{Escala}, {Gilbert}, {Wojno}, {Kirby}  \&
  {Guhathakurta}}{{Escala} et~al.}{2021}]{Escala2021}
{Escala} I.,  {Gilbert} K.~M.,  {Wojno} J.,  {Kirby} E.~N.,   {Guhathakurta}
  P.,  2021, \mn@doi [\aj] {10.3847/1538-3881/abfec4}, \href
  {https://ui.adsabs.harvard.edu/abs/2021AJ....162...45E} {162, 45}

\bibitem[\protect\citeauthoryear{{Escala}, {Gilbert}, {Fardal}, {Guhathakurta},
  {Sanderson}, {Kalirai}  \& {Mobasher}}{{Escala} et~al.}{2022}]{Escala2022}
{Escala} I.,  {Gilbert} K.~M.,  {Fardal} M.,  {Guhathakurta} P.,  {Sanderson}
  R.~E.,  {Kalirai} J.~S.,   {Mobasher} B.,  2022, \mn@doi [\aj]
  {10.3847/1538-3881/ac7146}, \href
  {https://ui.adsabs.harvard.edu/abs/2022AJ....164...20E} {164, 20}

\bibitem[\protect\citeauthoryear{{Fardal}, {Babul}, {Geehan}  \&
  {Guhathakurta}}{{Fardal} et~al.}{2006}]{Fardal2006}
{Fardal} M.~A.,  {Babul} A.,  {Geehan} J.~J.,   {Guhathakurta} P.,  2006,
  \mn@doi [\mnras] {10.1111/j.1365-2966.2005.09864.x}, \href
  {https://ui.adsabs.harvard.edu/abs/2006MNRAS.366.1012F} {366, 1012}

\bibitem[\protect\citeauthoryear{{Fardal}, {Guhathakurta}, {Babul}  \&
  {McConnachie}}{{Fardal} et~al.}{2007}]{Fardal2007}
{Fardal} M.~A.,  {Guhathakurta} P.,  {Babul} A.,   {McConnachie} A.~W.,  2007,
  \mn@doi [\mnras] {10.1111/j.1365-2966.2007.11929.x}, \href
  {https://ui.adsabs.harvard.edu/abs/2007MNRAS.380...15F} {380, 15}

\bibitem[\protect\citeauthoryear{{Fardal}, {Babul}, {Guhathakurta}, {Gilbert}
  \& {Dodge}}{{Fardal} et~al.}{2008}]{Fardal2008}
{Fardal} M.~A.,  {Babul} A.,  {Guhathakurta} P.,  {Gilbert} K.~M.,   {Dodge}
  C.,  2008, \mn@doi [\apjl] {10.1086/590386}, \href
  {https://ui.adsabs.harvard.edu/abs/2008ApJ...682L..33F} {682, L33}

\bibitem[\protect\citeauthoryear{{Fardal} et~al.,}{{Fardal}
  et~al.}{2012}]{Fardal2012}
{Fardal} M.~A.,  et~al., 2012, \mn@doi [\mnras]
  {10.1111/j.1365-2966.2012.21094.x}, \href
  {https://ui.adsabs.harvard.edu/abs/2012MNRAS.423.3134F} {423, 3134}

\bibitem[\protect\citeauthoryear{{Fardal} et~al.,}{{Fardal}
  et~al.}{2013}]{Fardal2013}
{Fardal} M.~A.,  et~al., 2013, \mn@doi [\mnras] {10.1093/mnras/stt1121}, \href
  {https://ui.adsabs.harvard.edu/abs/2013MNRAS.434.2779F} {434, 2779}

\bibitem[\protect\citeauthoryear{{Ferguson}, {Irwin}, {Ibata}, {Lewis}  \&
  {Tanvir}}{{Ferguson} et~al.}{2002}]{Ferguson2002}
{Ferguson} A. M.~N.,  {Irwin} M.~J.,  {Ibata} R.~A.,  {Lewis} G.~F.,   {Tanvir}
  N.~R.,  2002, \mn@doi [\aj] {10.1086/342019}, \href
  {https://ui.adsabs.harvard.edu/abs/2002AJ....124.1452F} {124, 1452}

\bibitem[\protect\citeauthoryear{{Ferguson}, {Johnson}, {Faria}, {Irwin},
  {Ibata}, {Johnston}, {Lewis}  \& {Tanvir}}{{Ferguson}
  et~al.}{2005}]{Ferguson2005}
{Ferguson} A. M.~N.,  {Johnson} R.~A.,  {Faria} D.~C.,  {Irwin} M.~J.,  {Ibata}
  R.~A.,  {Johnston} K.~V.,  {Lewis} G.~F.,   {Tanvir} N.~R.,  2005, \mn@doi
  [\apjl] {10.1086/429371}, \href
  {https://ui.adsabs.harvard.edu/abs/2005ApJ...622L.109F} {622, L109}

\bibitem[\protect\citeauthoryear{{Geehan}, {Fardal}, {Babul}  \&
  {Guhathakurta}}{{Geehan} et~al.}{2006a}]{Geehan2006}
{Geehan} J.~J.,  {Fardal} M.~A.,  {Babul} A.,   {Guhathakurta} P.,  2006a,
  \mn@doi [\mnras] {10.1111/j.1365-2966.2005.09863.x}, \href
  {https://ui.adsabs.harvard.edu/abs/2006MNRAS.366..996G} {366, 996}

\bibitem[\protect\citeauthoryear{{Geehan}, {Fardal}, {Babul}  \&
  {Guhathakurta}}{{Geehan} et~al.}{2006b}]{Geehan}
{Geehan} J.~J.,  {Fardal} M.~A.,  {Babul} A.,   {Guhathakurta} P.,  2006b,
  \mn@doi [\mnras] {10.1111/j.1365-2966.2005.09863.x}, \href
  {https://ui.adsabs.harvard.edu/abs/2006MNRAS.366..996G} {366, 996}

\bibitem[\protect\citeauthoryear{Gilbert, Fardal, Kalirai, Guhathakurta, Geha
  \& et al.}{Gilbert et~al.}{2007}]{Gilbert2007}
Gilbert K.~M.,  Fardal M.~A.,  Kalirai J.~S.,  Guhathakurta P.,  Geha M.~C.,
  et al. 2007, ApJ, 668, 245

\bibitem[\protect\citeauthoryear{Gilbert, Guhathakurta, Kollipara  \& et
  al.}{Gilbert et~al.}{2009}]{Gilbert2009}
Gilbert K.~M.,  Guhathakurta P.,  Kollipara P.,   et al. 2009, ApJ, 705, 1275

\bibitem[\protect\citeauthoryear{Gilbert, Kalirai, Guhathakurta  \& et
  al.}{Gilbert et~al.}{2014}]{Gilbert2014}
Gilbert K.~M.,  Kalirai J.~S.,  Guhathakurta P.,   et al. 2014, ApJ, 796, 76

\bibitem[\protect\citeauthoryear{Gilbert, Kirby, Escala  \& et al.}{Gilbert
  et~al.}{2019}]{Gilbert2019}
Gilbert K.~M.,  Kirby E.~N.,  Escala I.,   et al. 2019, ApJ, 883, 128

\bibitem[\protect\citeauthoryear{{Guhathakurta} et~al.,}{{Guhathakurta}
  et~al.}{2006}]{Guhathakurta}
{Guhathakurta} P.,  et~al., 2006, \mn@doi [\aj] {10.1086/499562}, \href
  {https://ui.adsabs.harvard.edu/abs/2006AJ....131.2497G} {131, 2497}

\bibitem[\protect\citeauthoryear{Hammer, Yang, Wang  \& et al.}{Hammer
  et~al.}{2010}]{Hammer2010}
Hammer F.,  Yang Y.~B.,  Wang J.~L.,   et al. 2010, ApJ, 725, 542

\bibitem[\protect\citeauthoryear{Hammer, Yang, Fouquet  \& et al.}{Hammer
  et~al.}{2013}]{Hammer2013}
Hammer F.,  Yang Y.~B.,  Fouquet S.,   et al. 2013, MNRAS, 431, 3543

\bibitem[\protect\citeauthoryear{Hammer, Yang, Wang, Ibata, Flores  \&
  Puech}{Hammer et~al.}{2018}]{Hammer2018}
Hammer F.,  Yang Y.~B.,  Wang J.~L.,  Ibata R.,  Flores H.,   Puech M.,  2018,
  MNRAS, 475, 2754

\bibitem[\protect\citeauthoryear{{Hernquist} \& {Quinn}}{{Hernquist} \&
  {Quinn}}{1988}]{Hernquist1988}
{Hernquist} L.,  {Quinn} P.~J.,  1988, \mn@doi [\apj] {10.1086/166592}, \href
  {https://ui.adsabs.harvard.edu/abs/1988ApJ...331..682H} {331, 682}

\bibitem[\protect\citeauthoryear{{Ibata}, {Irwin}, {Lewis}, {Ferguson}  \&
  {Tanvir}}{{Ibata} et~al.}{2001}]{Ibata2001}
{Ibata} R.,  {Irwin} M.,  {Lewis} G.,  {Ferguson} A. M.~N.,   {Tanvir} N.,
  2001, \nat, \href {https://ui.adsabs.harvard.edu/abs/2001Natur.412...49I}
  {412, 49}

\bibitem[\protect\citeauthoryear{{Ibata}, {Chapman}, {Ferguson}, {Irwin},
  {Lewis}  \& {McConnachie}}{{Ibata} et~al.}{2004}]{Ibata2004}
{Ibata} R.,  {Chapman} S.,  {Ferguson} A.~M.~N.,  {Irwin} M.,  {Lewis} G.,
  {McConnachie} A.,  2004, \mn@doi [\mnras] {10.1111/j.1365-2966.2004.07759.x},
  \href {https://ui.adsabs.harvard.edu/abs/2004MNRAS.351..117I} {351, 117}

\bibitem[\protect\citeauthoryear{{Ibata}, {Martin}, {Irwin}, {Chapman},
  {Ferguson}, {Lewis}  \& {McConnachie}}{{Ibata} et~al.}{2007}]{Ibata2007}
{Ibata} R.,  {Martin} N.~F.,  {Irwin} M.,  {Chapman} S.,  {Ferguson} A.~M.~N.,
  {Lewis} G.~F.,   {McConnachie} A.~W.,  2007, \mn@doi [\apj] {10.1086/522574},
  \href {https://ui.adsabs.harvard.edu/abs/2007ApJ...671.1591I} {671, 1591}

\bibitem[\protect\citeauthoryear{{Johnston}, {Sackett}  \&
  {Bullock}}{{Johnston} et~al.}{2001}]{Johnston2001}
{Johnston} K.~V.,  {Sackett} P.~D.,   {Bullock} J.~S.,  2001, \mn@doi [\apj]
  {10.1086/321644}, \href
  {https://ui.adsabs.harvard.edu/abs/2001ApJ...557..137J} {557, 137}

\bibitem[\protect\citeauthoryear{{Johnston}, {Bullock}, {Sharma}, {Font},
  {Robertson}  \& {Leitner}}{{Johnston} et~al.}{2008}]{Johnston2008}
{Johnston} K.~V.,  {Bullock} J.~S.,  {Sharma} S.,  {Font} A.,  {Robertson}
  B.~E.,   {Leitner} S.~N.,  2008, \mn@doi [\apj] {10.1086/592228}, \href
  {https://ui.adsabs.harvard.edu/abs/2008ApJ...689..936J} {689, 936}

\bibitem[\protect\citeauthoryear{{Kacharov} et~al.,}{{Kacharov}
  et~al.}{2017}]{Kacharov}
{Kacharov} N.,  et~al., 2017, \mn@doi [\mnras] {10.1093/mnras/stw3188}, \href
  {https://ui.adsabs.harvard.edu/abs/2017MNRAS.466.2006K} {466, 2006}

\bibitem[\protect\citeauthoryear{Kalirai, Guhathakurta, Gilbert  \& et
  al.}{Kalirai et~al.}{2006}]{Kalirai2006a}
Kalirai J.~S.,  Guhathakurta P.,  Gilbert K.~M.,   et al. 2006, ApJ, 641, 268

\bibitem[\protect\citeauthoryear{{Karademir}, {Remus}, {Burkert}, {Dolag},
  {Hoffmann}, {Moster}, {Steinwandel}  \& {Zhang}}{{Karademir}
  et~al.}{2019}]{Karademir}
{Karademir} G.~S.,  {Remus} R.-S.,  {Burkert} A.,  {Dolag} K.,  {Hoffmann}
  T.~L.,  {Moster} B.~P.,  {Steinwandel} U.~P.,   {Zhang} J.,  2019, \mn@doi
  [\mnras] {10.1093/mnras/stz1251}, \href
  {https://ui.adsabs.harvard.edu/abs/2019MNRAS.487..318K} {487, 318}

\bibitem[\protect\citeauthoryear{Kirihara, Miki  \& Mori}{Kirihara
  et~al.}{2017}]{Kirihara2017}
Kirihara T.,  Miki Y.,   Mori M. e.~a.,  2017, MNRAS, 464, 3509

\bibitem[\protect\citeauthoryear{Koleva, Prugniel, De~Rijcke, Zeilinger  \&
  Michielsen}{Koleva et~al.}{2009a}]{Koleva2009b}
Koleva M.,  Prugniel P.,  De~Rijcke S.,  Zeilinger W.~W.,   Michielsen D.,
  2009a, AN, 330, 960

\bibitem[\protect\citeauthoryear{Koleva, de Rijcke, Prugniel, Zeilinger  \&
  Michielsen}{Koleva et~al.}{2009b}]{Koleva2009a}
Koleva M.,  de Rijcke S.,  Prugniel P.,  Zeilinger W.~W.,   Michielsen D.,
  2009b, MNRAS, 396, 2133

\bibitem[\protect\citeauthoryear{{Leaman} et~al.,}{{Leaman}
  et~al.}{2013}]{Leaman}
{Leaman} R.,  et~al., 2013, \mn@doi [\apj] {10.1088/0004-637X/767/2/131}, \href
  {https://ui.adsabs.harvard.edu/abs/2013ApJ...767..131L} {767, 131}

\bibitem[\protect\citeauthoryear{{McConnachie}, {Irwin}, {Ibata}, {Ferguson},
  {Lewis}  \& {Tanvir}}{{McConnachie} et~al.}{2003}]{McConnachie2003}
{McConnachie} A.~W.,  {Irwin} M.~J.,  {Ibata} R.~A.,  {Ferguson} A.~M.~N.,
  {Lewis} G.~F.,   {Tanvir} N.,  2003, \mn@doi [\mnras]
  {10.1046/j.1365-8711.2003.06785.x}, \href
  {https://ui.adsabs.harvard.edu/abs/2003MNRAS.343.1335M} {343, 1335}

\bibitem[\protect\citeauthoryear{{Mercado} et~al.,}{{Mercado}
  et~al.}{2021}]{Mercado}
{Mercado} F.~J.,  et~al., 2021, \mn@doi [\mnras] {10.1093/mnras/staa3958},
  \href {https://ui.adsabs.harvard.edu/abs/2021MNRAS.501.5121M} {501, 5121}

\bibitem[\protect\citeauthoryear{{Merrett} et~al.,}{{Merrett}
  et~al.}{2006}]{Merret2006}
{Merrett} H.~R.,  et~al., 2006, \mn@doi [\mnras]
  {10.1111/j.1365-2966.2006.10268.x}, \href
  {https://ui.adsabs.harvard.edu/abs/2006MNRAS.369..120M} {369, 120}

\bibitem[\protect\citeauthoryear{{Miki}, {Mori}  \& {Rich}}{{Miki}
  et~al.}{2016}]{MikiMori}
{Miki} Y.,  {Mori} M.,   {Rich} R.~M.,  2016, \mn@doi [\apj]
  {10.3847/0004-637X/827/1/82}, \href
  {https://ui.adsabs.harvard.edu/abs/2016ApJ...827...82M} {827, 82}

\bibitem[\protect\citeauthoryear{{Milo{\v{s}}evi{\'c}}}{{Milo{\v{s}}evi{\'c}}}{2022}]{Milosevic2022}
{Milo{\v{s}}evi{\'c}} S.,  2022, \mn@doi [Serbian Astronomical Journal] {DOI
  10.2298/SAJ220704004M}, \href
  {https://ui.adsabs.harvard.edu/abs/2022SerAJ.tmp....4M} {}

\bibitem[\protect\citeauthoryear{{Milo{\v{s}}evi{\'c}}, {Mi{\'c}i{\'c}}  \&
  {Lewis}}{{Milo{\v{s}}evi{\'c}} et~al.}{2022}]{Milosevic}
{Milo{\v{s}}evi{\'c}} S.,  {Mi{\'c}i{\'c}} M.,   {Lewis} G.~F.,  2022, \mn@doi
  [\mnras] {10.1093/mnras/stac249}, \href
  {https://ui.adsabs.harvard.edu/abs/2022MNRAS.511.2868M} {511, 2868}

\bibitem[\protect\citeauthoryear{{Mori} \& {Rich}}{{Mori} \&
  {Rich}}{2008}]{MoriRich}
{Mori} M.,  {Rich} R.~M.,  2008, \mn@doi [\apjl] {10.1086/529140}, \href
  {https://ui.adsabs.harvard.edu/abs/2008ApJ...674L..77M} {674, L77}

\bibitem[\protect\citeauthoryear{{Navarro}, {Frenk}  \& {White}}{{Navarro}
  et~al.}{1996}]{Navarro1996}
{Navarro} J.~F.,  {Frenk} C.~S.,   {White} S. D.~M.,  1996, \mn@doi [\apj]
  {10.1086/177173}, \href
  {https://ui.adsabs.harvard.edu/abs/1996ApJ...462..563N} {462, 563}

\bibitem[\protect\citeauthoryear{{Pillepich} et~al.,}{{Pillepich}
  et~al.}{2014}]{Pillepich}
{Pillepich} A.,  et~al., 2014, \mn@doi [\mnras] {10.1093/mnras/stu1408}, \href
  {https://ui.adsabs.harvard.edu/abs/2014MNRAS.444..237P} {444, 237}

\bibitem[\protect\citeauthoryear{Remus, Dolag, Naab, Burkert, Hirschmann,
  Hoffmann  \& Johansson}{Remus et~al.}{2016}]{Remus}
Remus R.-S.,  Dolag K.,  Naab T.,  Burkert A.,  Hirschmann M.,  Hoffmann T.~L.,
    Johansson P.~H.,  2016, \mn@doi [Monthly Notices of the Royal Astronomical
  Society] {10.1093/mnras/stw2594}, 464, 3742

\bibitem[\protect\citeauthoryear{{Richardson} et~al.,}{{Richardson}
  et~al.}{2008}]{Richardson2008}
{Richardson} J.~C.,  et~al., 2008, \mn@doi [\aj]
  {10.1088/0004-6256/135/6/1998}, \href
  {https://ui.adsabs.harvard.edu/abs/2008AJ....135.1998R} {135, 1998}

\bibitem[\protect\citeauthoryear{{Sadoun}, {Mohayaee}  \& {Colin}}{{Sadoun}
  et~al.}{2014}]{Sadoun}
{Sadoun} R.,  {Mohayaee} R.,   {Colin} J.,  2014, \mn@doi [\mnras]
  {10.1093/mnras/stu850}, \href
  {https://ui.adsabs.harvard.edu/abs/2014MNRAS.442..160S} {442, 160}

\bibitem[\protect\citeauthoryear{Spolaor, Proctor, Forbes  \& Couch}{Spolaor
  et~al.}{2009}]{Spolaor}
Spolaor M.,  Proctor R.~N.,  Forbes D.~A.,   Couch W.~J.,  2009, ApJL, 691,
  L138

\bibitem[\protect\citeauthoryear{{Springel}}{{Springel}}{2005}]{Springel}
{Springel} V.,  2005, \mn@doi [\mnras] {10.1111/j.1365-2966.2005.09655.x},
  \href {https://ui.adsabs.harvard.edu/abs/2005MNRAS.364.1105S} {364, 1105}

\bibitem[\protect\citeauthoryear{{Tanaka}, {Chiba}, {Komiyama}, {Guhathakurta},
  {Kalirai}  \& {Iye}}{{Tanaka} et~al.}{2010}]{Tanaka2010}
{Tanaka} M.,  {Chiba} M.,  {Komiyama} Y.,  {Guhathakurta} P.,  {Kalirai} J.~S.,
    {Iye} M.,  2010, \mn@doi [\apj] {10.1088/0004-637X/708/2/1168}, \href
  {https://ui.adsabs.harvard.edu/abs/2010ApJ...708.1168T} {708, 1168}

\bibitem[\protect\citeauthoryear{{White} \& {Frenk}}{{White} \&
  {Frenk}}{1991}]{White2}
{White} S. D.~M.,  {Frenk} C.~S.,  1991, \mn@doi [\apj] {10.1086/170483}, \href
  {https://ui.adsabs.harvard.edu/abs/1991ApJ...379...52W} {379, 52}

\bibitem[\protect\citeauthoryear{{White} \& {Rees}}{{White} \&
  {Rees}}{1978}]{White1}
{White} S.~D.~M.,  {Rees} M.~J.,  1978, \mn@doi [\mnras]
  {10.1093/mnras/183.3.341}, \href
  {https://ui.adsabs.harvard.edu/abs/1978MNRAS.183..341W} {183, 341}

\bibitem[\protect\citeauthoryear{{Widrow}, {Pym}  \& {Dubinski}}{{Widrow}
  et~al.}{2008}]{Widrow}
{Widrow} L.~M.,  {Pym} B.,   {Dubinski} J.,  2008, \mn@doi [\apj]
  {10.1086/587636}, \href
  {https://ui.adsabs.harvard.edu/abs/2008ApJ...679.1239W} {679, 1239}

\bibitem[\protect\citeauthoryear{{Wojno} et~al.,}{{Wojno} et~al.}{2023}]{Wojno}
{Wojno} J.~L.,  et~al., 2023, \mn@doi [\apj] {10.3847/1538-4357/acd5d3}, \href
  {https://ui.adsabs.harvard.edu/abs/2023ApJ...951...12W} {951, 12}

\makeatother
\end{thebibliography}

%\begin{thebibliography}{99}
%\end{thebibliography}

%%%%%%%%%%%%%%%%%%%%%%%%%%%%%%%%%%%%%%%%%%%%%%%%%%

%%%%%%%%%%%%%%%%% APPENDICES %%%%%%%%%%%%%%%%%%%%%

% Don't change these lines
%\bsp	% typesetting comment
\label{lastpage}
\end{document}